# Spin exchange of two spin-1/2 atoms*

PAN Zeming, TAN Naiming, GAO Chao, YAO Zhihai, WANG Xiaoqian

Department of Physics, Changchun University of Science and Technology, Changchun 130022, China

**Abstract**

The quantum Cheshire cat effect is an important phenomenon in quantum mechanics that reveals the separability of physical properties from their carriers. This effect transcends the classical framework whose attributes must be inherently attached to objects, providing new perspectives for quantum information and precision measurement. According to the quantum Cheshire cat effect, we prepare a pre-selected state of a spin-1/2 atomic system composed of two particles through a pre-selection process. We conduct quantum weak measurements on the spins and positions of these two atoms and extract weak values by using the method of imaginary time evolution (ITE). Subsequently, we perform post-selection on these two atoms and design two distinct post-selected states. Initially, we calculate analytical solutions when both atoms encounter these two different post-selected states separately. Then, during the stage of obtaining weak values via ITE, we first discuss the scenario with only one post-selected state. In this case, our experimental goal is to achieve spin exchange between the two atoms. We use ITE to obtain normalized coincidence rate for the system. By linearly fitting these normalized coincidence rate, we derive numerical solutions for the weak values of the system. The comparison between the analytical solutions and numerical results indicates that they are in close agreement, demonstrating that our method promotes spin exchange between the two atoms. Next, we examine scenarios involving both post-selected states in the post-selection process. After completing weak measurements on particles, delayed-choice allows them to evolve along different paths ultimately leading to distinct post-selected states. One particular post-selected state that results in final measurement outcomes indicates that the spin exchange occurs between both particles with amplification. Conversely, the other post-selected state ensures that even after undergoing weak measurement and delayed-choice, the states of the two particles remain consistent with their pre-measurement conditions. We also compare the analytical and numerical solutions of the experiment involving delayed choice and find that they are very consistent with each other. This consistency indicates that delayed-choice indeed has a significant influence on whether the final exchange occurs. Our research

---

theoretically confirms the feasibility of fermionic systems within bipartite quantum Cheshire cat effects and illustrates how delayed-choice influences quantum Cheshire cat effects in spin-1/2 atomic systems.

Keywords: quantum information，weak value，quantum Cheshire cat

**PACS:** 03.65.Ta, 03.67.–a, 03.67.Ac

# 1. Introduction

The quantum Cheshire cat has attracted wide attention because it reveals the flexible relationship between matter and its physical properties in the quantum world. As early as 2013, researchers discovered the phenomenon that the polarization of photons can be separated from the body[1] on the basis of weak measurement theory, and named the phenomenon that a certain property of an object is separated from the body as a quantum Cheshire cat. Based on this phenomenon, researchers have realized a variety of quantum Cheshire Cat experiments, which includes microscopic particles such as photons, electrons and neutrons[2-12]. The weak measurement techniques that can realize the quantum Cheshire cat phenomenon can be divided into two categories[13]: one is to obtain the relevant information of the system indirectly through the direct measurement of the pointer by using the weak coupling[13,14] between the system and the pointer; The other is that through the weak coupling between the system and the environment, the imaginary time evolution (ITE)[7,12,17] is carried out by means of the optical lattice[15] and the particle interferometer[16], the relationship between the environment and the system is established according to the imaginary time, and the relevant information of the system is obtained through the disturbance of the environment in the imaginary time. For the first weak measurement method mentioned earlier, because the time-evolving mechanical quantity to be constructed is related to both the pointer and the system, the direct measurement of the wave function of the pointer will not cause the collapse of the wave function of the system, and the coupling between the two can be simply constructed, so it is mostly used to measure a single particle. In a system of two or more particles, because there are too many targets to be measured, the introduction of pointers will lead to extremely complex measurement and calculation, while ITE does not introduce additional pointers, and can maintain the quantum coherence of the system to the greatest extent, so ITE has greater advantages in dealing with many-body wave functions or multi-qubit wave functions.

So far, the quantum Cheshire cat effect has been widely studied in single-particle systems. For example, in a single-particle system, the separation[4] of the position and spin of a photon is realized by non-invasive weak measurement; the separation[6] of the position and spin of a photon is realized under the condition of decoherence; the system is separated in a chaotic environment, and the amplification[9] of an optical signal is realized; after the body and polarization of a photon are separated, the polarization is changed by weak measurement in a space without a photon for[10]; Realize the separation of the wave and particle properties of photons, i.e., measuring the wave and particle characteristics of photons on separate paths [7],Send neutrons into a silicon crystal interferometer and perform weak measurements on their position and magnetic moment to achieve the separation of the position and spin of neutrons [2],Use a crystal interferometer to achieve the separation of the position, spin, and energy attributes of neutrons [3]. In many-particle systems, [11] and [12] constructed spin-dependent and position-dependent pre-selected States based on the exchange symmetry of the wave function of the photon pair, and realized the permanent exchange of the spin of two photons through the quantum Cheshire cat effect. At present, the study of the quantum Cheshire cat effect in a many-fermion system is not perfect, so in this paper, we take spin-1/2 as an example, construct the pre-selected States of the two with respect to position and spin, and perform weak measurement on the two pre-selected States and extract the weak value, thus completing the numerical experiment of the quantum Cheshire cat effect in a many-fermion system.

The structure of this paper is as follows: Section 2 introduces the basic concept of weak value and the theoretical framework of spin-1/2 atomic pairs in the quantum Cheshire cat effect; In section 3, the weak measurement is performed on the pre-selected state of two spin-1/2 atom pairs, and then the weak measurement is performed on the atom pair after the delayed selection is introduced; Conclusions are given in Section 4.

**2. Theoretical framework**

Aharonov et al.[13] introduced a physical quantity called weak value, and found in experiments through the two-state vector fomalism[18] that the observable of the particle spin component can be amplified from 1/2 to 100 by the weak value, that is, the weak value can increase the observable. It can be seen that the weak value is different from the expectation value obtained by the traditional direct measurement. It is a value determined by two different initial and final state vectors and the weak measurement. Assuming that $\hat{A}$ is any observable in the system, the weak value $\langle A \rangle_w$ with respect to $\hat{A}$ can be written as:

$$\langle A \rangle_w = \frac{\langle f|\hat{A}|i\rangle}{\langle f|i\rangle}. \qquad (1)$$

Where $|i\rangle$ is the pre-selected state of the system, and $|f\rangle$ is the post-selected state of the system. The $|i\rangle$ and $|f\rangle$ can be set artificially according to the measured mechanical quantities. For a spin-1/2 atomic system, we discuss the weak measurement of the position and spin of the atom. The measurement operators with respect to the position are $\Pi_u = |u\rangle\langle u|$ and $\Pi_d = |d\rangle\langle d|$ (the subscript u represents the up path and d represents the down path). The measurement operator of the spin is $S_z = 1/2(|\uparrow\rangle\langle\uparrow| - |\downarrow\rangle\langle\downarrow|)$ ($\hbar=1$), where $|\uparrow\rangle$ and $|\downarrow\rangle$, $|u\rangle$ and $|d\rangle$ are orthogonal to each other, and $u$ and $d$ are the two possible paths for the atom. In addition, the quantum Cheshire cat effect and the delayed choice experiment[19] proposed by Wheeler both embody the principle of time symmetry in quantum mechanics, that is, the properties of a system are affected not only by past events, but also by future events. Reference [5] and Reference [8] combine the quantum Cheshire cat effect and delayed selection, and realize the delayed selection of single photon and single neutron under the quantum Cheshire cat effect, as well as the bulk and spin separation of photon and neutron, respectively. Therefore, in this paper, we introduce the delayed selection into the Cheshire cat experiment while studying the spin exchange of the Fermion multibody system, and then discuss the results of the interaction between the delayed selection and the quantum Cheshire cat effect in the Fermion multibody system.

The experimental principle is shown in Fig. 1. In this paper, two spin-1/2 atoms are considered, and the spins of the two atoms are exchanged by preparing appropriate pre-selected States and post-selected States. In the pre-selected part, the pre-selected States of the two particles are assumed to be linear cluster States[13,20-22],

$$|i\rangle = \frac{1}{\sqrt{2}}(|\uparrow_1\downarrow_2\rangle - |\downarrow_1\uparrow_2\rangle)(sin\alpha|u_1 d_2\rangle + cos\alpha|d_1 u_2\rangle) \qquad (2)$$

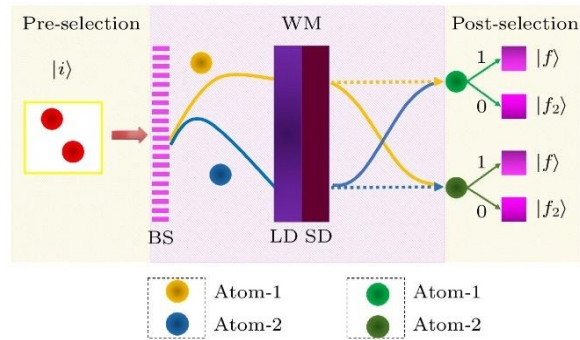

**Figure 1.** Schematic diagram of spin-1/2 atomic spin exchange and spin amplification principle. In the pre-selection section, prepare a pre-selected state $|i\rangle$ that meets theoretical expectations. In the weak measurement (WM) section, the beam splitter (BS) splits two atoms into beams, which then enter the position density processor (LD) and spin-sensitive density processor (SD). In the post-selection section, two atoms will select synchronlusly through a random switch controlled by 1 and 0, ensuring two atoms simultaneously randomly obtain one of the post-selected states. After two atoms pass through BS, the atom passing through the downward path is called atom 1, represented in yellow; the atom passing through the upper

path is called atom 2, represented in blue. After completing the weak measurement, use light green to represent atom 1 and dark green to represent atom 2.

The two particles in this state are subjected to a weak measurement. First, the two particles are divided into two beams by a beam splitter. In this paper, the atom passing through the lower path of Fig. 1 after beam splitting is called atom 1, and the atom passing through the upper path is called atom 2. The two particles after beam splitting enter the position density processor and the spin sensitive density processor respectively, and the two processors are used to simulate the imaginary time perturbation of the position and spin of the particles respectively. After the perturbation, the two particles will enter the post-selection stage. In the post-selection stage, the two particles face a two ways switch simultaneously and synchronously, which is controlled by a random number generator that generates 1 or 0 randomly in real time. When the two particles come to the switch, if the number generated by the generator at this moment is 1, the two particles will obtain the post-selected state $|f\rangle$ in the post-selection,

$$|f\rangle = \frac{1}{2}(|\uparrow_1\downarrow_2\rangle + |\downarrow_1\uparrow_2\rangle)|u_1 d_2\rangle + \frac{1}{2}(|\uparrow_1\downarrow_2\rangle - |\downarrow_1\uparrow_2\rangle)|d_1 u_2\rangle. \quad (3)$$

When two particles arrive at the switch and the number generated by the generator is 0, the post-selected state obtained is

$$|f\rangle = |\uparrow_1\downarrow_2\rangle|d_1 u_2\rangle. \quad (4)$$

When the pre-selected state $|i\rangle$ (2) and the post-selected state $|f\rangle$ (3) and the positions and spins of the two particles are substituted into the expression of the weak value (1), the analytical solution of the two atoms with respect to the positions and spins when the two atoms exchange spins is obtained:

$$\langle \Pi_u^1 \rangle_w = 0, \ \langle \Pi_u^2 \rangle_w = 1, \ \langle \Pi_d^1 \rangle_w = 1, \ \langle \Pi_d^2 \rangle_w = 0. \quad (5)$$

$$\langle \Pi_u^1 \otimes S_1 \rangle_w = \frac{1}{2}\tan\alpha, \ \langle \Pi_u^2 \otimes S_2 \rangle_w = 0,$$
$$\langle \Pi_d^1 \otimes S_1 \rangle_w = 0, \ \langle \Pi_d^2 \otimes S_2 \rangle_w = -\frac{1}{2}\tan\alpha. \quad (6)$$

The non-zero weak value of the observed value indicates that the system is indeed in the pre-selected state and the post-selected state. On the contrary, when the weak value is zero, it indicates that the system is not in the state. According to the results of the analytical solution, it can be found that atom 1 can indeed appear in the lower path and atom 2 can appear in the upper path by selecting the pre-selected state and the post-selected state, and the spin of atom 2 can be observed in the lower path and the spin of atom 1 can be observed in the upper path. $\langle \Pi_d^1 \otimes S_1 \rangle_w$ and $\langle \Pi_u^2 \otimes S_2 \rangle_w$ are tan functions with respect to the $\alpha$. This shows that when the regulatory parameter $\alpha$ approaches 0, that is, $|\langle f|i \rangle|^2$ approaches 1, the measurement success probability is very large; When the $\alpha$ approaches the $\pi/2$, the weak value amplification can be realized. The value of $\alpha$ can be set according to different situations to obtain the expected experimental results. Only the value of $\alpha$ in one period needs to be considered, that is, the value range is $(0,\pi/2)$. Because when tan $\alpha$ is not 0, the spin of atom 1 can be found on the u path and the spin of atom 2 can be found on the d path. It means that the spins of the two atoms have been exchanged. If tan $\alpha$ is 0, it means that the pre-selected state of the two particles becomes $|i_{\tan\alpha=0}\rangle = 1/\sqrt{2}(|\uparrow_1\downarrow_2\rangle - |\downarrow_1\uparrow_2\rangle)|d_1u_2\rangle$, and the state vector after the spin operator acts on the pre-selected state will be completely orthogonal to the post-selected state, resulting in the weak value always being 0, so it is not necessary to consider the case where $\alpha$ is 0. Similarly, when $\alpha$ is $\pi/2$, the measurement probability, that is, $|\langle f|i_{\tan\alpha=\pi/2}\rangle|^2$, is 0, so no valid measurement result will be obtained, and this situation does not need to be considered. When the pre-selected state $|i\rangle$ (2) and the post-selected state $|f_2\rangle$ (4) and the mechanical quantities of the position and spin of the two particles are substituted into the expression of the weak value (1), the analytical solution of the position and spin of the two atoms is obtained when the two atoms do not exchange spins. It should also be noted that the reason why the spin weak value is not measured separately in this study is that the spins of the two atoms must exist and there is no need to measure separately.

Similarly, according to the weak value formula, the analytical solution of the system without commutation is obtained as

$$\langle \Pi_u^1 \rangle_w = 0, \quad \langle \Pi_u^2 \rangle_w = 1, \quad \langle \Pi_d^1 \rangle_w = 1, \quad \langle \Pi_d^2 \rangle_w = 0. \tag{7}$$

$$\begin{aligned} \langle \Pi_u^1 \otimes S_1 \rangle_{w2} &= 0, \quad \langle \Pi_u^2 \otimes S_2 \rangle_{w3} = 0.5, \\ \langle \Pi_d^1 \otimes S_1 \rangle_{w2} &= -0.5, \quad \langle \Pi_d^2 \otimes S_1 \rangle_{w3} = 0. \end{aligned} \tag{8}$$

The weak value results show that the spins of the two particles are not exchanged, but the results are slightly different from the eigenvalues due to the influence of the initial and final state vectors, and the spin of atom 2 is negative. The two different post-selected States lead to different measurement results, which fully reflects the time symmetry in the quantum world. Even if the weak measurement has been completed, the delayed selection that occurs later also affects the previous measurement.

### 3. Extraction of weak values by ITE

Section [2] presents the results of the analytical solutions for the presets and weak values of the pre-selected and post-selected States. Next, we will describe in detail how to obtain the numerical solution of the experiment by means of ITE. ITE originated from the[23] of Wick rotation in special relativity, and now it is mostly used in different fields such as quantum field theory,[24] and quantum simulation[25]. ITE is a weak value extraction method based on the disturbance of the system without introducing any auxiliary pointer state. Non-unitary matrix generated by the observable quantities $\Pi$ or $\Pi \otimes S$ of the two particles:

$$U(H,t) = e^{-Ht} \tag{9}$$

The role of $U(H,t)$ is to evolve the system in imaginary time[26], where $H$ is the Hamiltonian of the system, the role of $H$ is to generate the operator of the evolution of the system, the expression of $H$ is $H=O$, and $O$ represents the observable. The parameter $t$ is a time parameter in imaginary form and is generally not restricted. However, in ITE, the $t$ must be small enough to minimize the disturbance to the system during the interaction time. The advantage of ITE to extract the weak value is that there is no need to introduce an additional parameter state, which greatly simplifies the problem, and the weak value can still be extracted by ITE without involving the imaginary number, in which case the real part of the weak value of the Hamiltonian is proportional to the slope of the function obtained by ITE. In ITE, the probabilities of successful post-selection before and after applying the perturbation are $N_0 = |\langle f|i\rangle|^2$ and $N(U) = |\langle f|U|i\rangle|^2$. The normalized coincidence rate $N(t)$ is related to the former two by

$$N(t) = \frac{N(U)}{N_0}. \tag{10}$$

In this experiment, the weak value as a function of $N(t)$ is

$$-\frac{1}{2}\frac{\partial N}{\partial t}\bigg|_{t\to 0} = Re\langle O\rangle_w. \tag{11}$$

First, the pre-selected state and post-selected state of the system to be measured are determined to obtain the $N_0$, and then the normalized coincidence rate of the position and spin of the two atoms can be obtained by substituting the mechanical quantities of the position and spin of the two atoms into the $N(U)$. In order to simulate the perturbation of the environment, the weak value of the system is obtained after the imaginary time evolution $N(t)$. Because the position density processor and the spin sensitive density processor are important experimental devices for weak measurement to obtain the position weak value of the system. Therefore, in the weak measurement part of this paper, it is assumed that a series of position density processors with different transmissivities (the relationship between the transmissivity $T$ and the $t$ is $T = e^{-2t}$) are used to simulate the perturbation of position and spin. According to the evolution of imaginary time, the transmittance also changes with it, and the normalized coincidence rate of the system under different imaginary time is calculated. The curve is obtained by linear fitting of these imaginary time-dependent data points, and the numerical solution of the weak value is obtained by analyzing the curve according to (11).

First, we discuss the case where delayed selection is not considered, that is, the post-selected state of the system must be (3).

As shown in the Fig. 2, when the post-selected state is (3), different weak value results can be obtained by adjusting the $\alpha$ of the pre-selected state. Substituting $\Pi_u^1$, $\Pi_d^2$, $\Pi_u^2$, $\Pi_d^1$, $\Pi_d^1 \otimes S_1$, $\Pi_u^2 \otimes S_2$, $\Pi_d^1 \otimes S_1$ and $\Pi_u^2 \otimes S_2$ into (9) respectively to obtain the related non-unitary matrices, and th/en substituting these non-unitary matrices and a variety of different pre-selected States obtained by adjusting the value of $\alpha$ into (10) to obtain the normalized coincidence rate of each mechanical quantity under the imaginary time evolution. In one period, i.e. $\alpha \in (-\pi/2, \pi/2)$, the normalized coincidence rates for the observables $\Pi_u^1$, $\Pi_d^2$, $\Pi_d^1 \otimes S_1$ and $\Pi_u^2 \otimes S_2$ are also almost unaffected by $t$ and $\alpha$. The change of the normalized coincidence rate of $\Pi_u^2$ and $\Pi_d^1$ is only related to $t$. The normalized coincidence rate of $\Pi_d^1 \otimes S_1$ and $\Pi_u^2 \otimes S_2$ has a certain functional relationship with $t$ and $\alpha$. Combining the analytical solutions of (5) and (6), it can be concluded that the weak value results of $\Pi_u^1$ and $\Pi_d^2$ are theoretically independent of $\alpha$, but the normalized coincidence rate of the experimental results decreases with the decrease of the value of $\alpha$. This shows that the small interaction of the environment on the system may have an impact on the system. With the time evolution, the parameters in the pre-selected state are not completely eliminated due to the orthogonality of the States, indicating that the entanglement of the pre-selected state is changed by the environment. For $\Pi_d^1 \otimes S_1$ and $\Pi_u^2 \otimes S_2$, $N$ has the same trend with $\alpha$, but has the opposite trend with $t$.

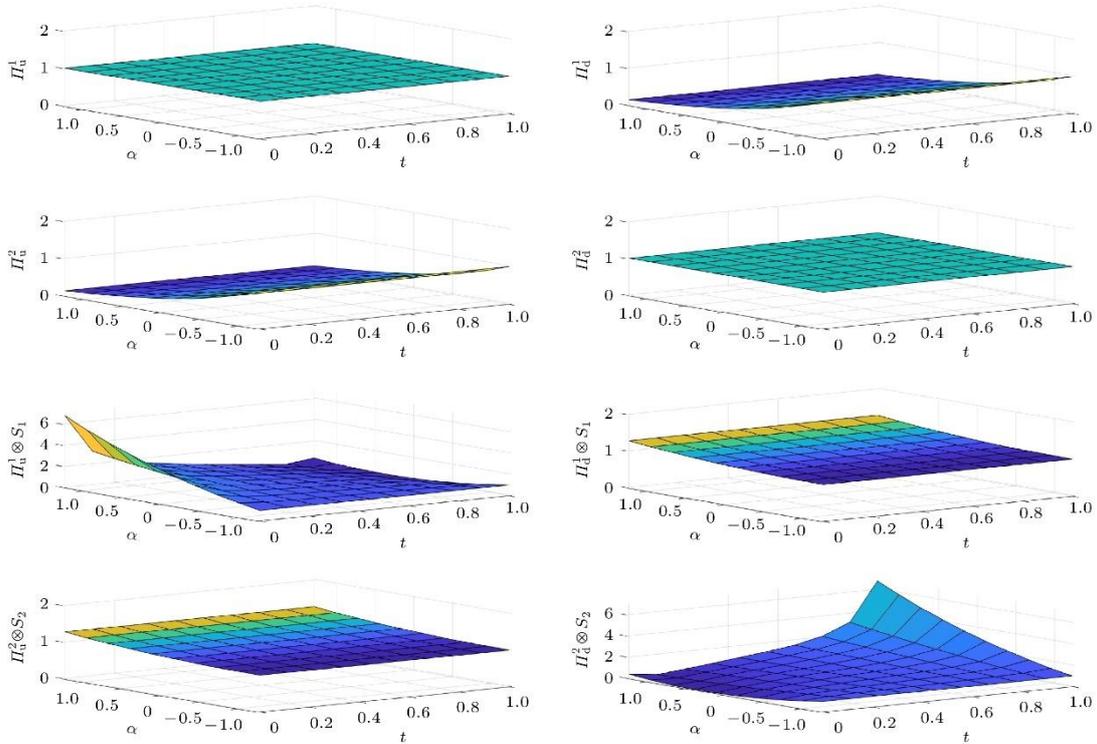

**Figure 2.** Trend of normalized coincidence rate $N(t)$ as a function of $t$ and $\alpha$. The value range

of $t$ in the figure is from 0 to 1, and the value range of $\alpha$ in the figure is from 0 to $\pi/2$.

In order to simplify the extraction of the weak value by the normalized coincidence rate, the $\alpha = \pi/4$ is taken as an example to extract the weak value. In this case, the pre-selected state of the system is

$$|i_1\rangle = \frac{1}{2}(|\uparrow_1\downarrow_2\rangle - |\downarrow_1\uparrow_2\rangle)(|u_1 d_2\rangle + |d_1 u_2\rangle). \tag{12}$$

The analytical solution of the system is obtained by the pre-selected state (12) and the post-selected state (3).

$$\langle \Pi_u^1 \rangle_w = 0, \quad \langle \Pi_u^2 \rangle_w = 1, \quad \langle \Pi_d^1 \rangle_w = 1, \quad \langle \Pi_d^2 \rangle_w = 0. \tag{13}$$

$$\begin{aligned} \langle \Pi_u^1 \otimes S_1 \rangle_w &= 0.5, & \langle \Pi_u^2 \otimes S_2 \rangle_w &= 0, \\ \langle \Pi_d^1 \otimes S_1 \rangle_w &= 0, & \langle \Pi_d^2 \otimes S_2 \rangle_w &= -0.5. \end{aligned} \tag{14}$$

Then, the numerical solutions and analytical solutions of the experiment are compared through ITE extraction. When $\alpha$ is set to $\pi/4$, the method for obtaining the normalized coincidence rate of each mechanical quantity is the same as that for the pre-selected states when $\alpha$ varies.

As shown in Fig. 3, where the normalized coincidence rate $N(t) = \frac{N(U)}{N_0}$ is the ordinate of the sampled data points, and the time $t$ is the abscissa, the numerical solution can be obtained from the straight line fitted from the normalized data points. Data on the $N(t)$ of the positions and spins of the two atoms obtained after the ITE of the system. The normalized coincidence rates of $\langle \Pi_u^1 \rangle_w$, $\langle \Pi_d^2 \rangle_w$, $\langle \Pi_u^2 \otimes S_2 \rangle_w$ and $\langle \Pi_d^1 \otimes S_1 \rangle_w$ have low correlation with the interaction time $t$, while the normalized coincidence rates of $\langle \Pi_u^2 \rangle_w$, $\langle \Pi_d^1 \rangle_w$, $\langle \Pi_u^1 \otimes S_1 \rangle_w$ and $\langle \Pi_d^2 \otimes S_2 \rangle_w$ have obvious functional relationship with the interaction time $t$. The weak values are obtained by linear fitting and probability correction of 8 curves in Fig. 3. The normalized coincidence rate of each observation is obtained by formula (10), and the normalized coincidence rate of each observation is substituted into formula (11),

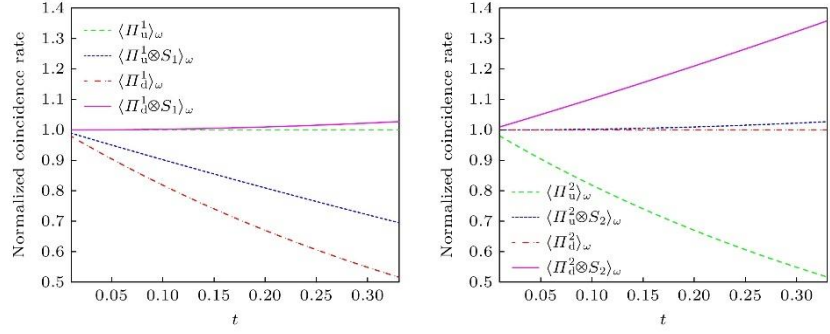

**Figure 3.** When α takes π/4, the normalized coincidence rate $N(t)$ of two atoms varies with $t$. The left image shows the data for atom 1, and the right image shows the data for atom 2. Due to the functional relationship between $N(t)$ and $t$, weak values are directly used here to represent the $N(t)$ of the relevant observables.

$$\langle \Pi_u^1 \rangle_w = 0.00, \quad \langle \Pi_u^2 \rangle_w = 0.84, \\ \langle \Pi_d^1 \rangle_w = 0.82, \quad \langle \Pi_d^2 \rangle_w = 0.00. \tag{15}$$

$$\langle \Pi_u^1 \otimes S_1 \rangle_w = 0.48, \quad \langle \Pi_u^2 \otimes S_2 \rangle_w = -0.03, \\ \langle \Pi_d^1 \otimes S_1 \rangle_w = -0.05, \quad \langle \Pi_d^2 \otimes S_1 \rangle_w = -0.53. \tag{16}$$

The numerical solution of the weak value extracted by ITE is then obtained. For the position and spin entangled state of atom 1 and atom 2, it is almost impossible to determine the position of the two atoms and the spin state of the two atoms by direct measurement of their wave functions. Multiple measurements will give a probability of 1/2 for both atoms to appear in the u and d paths and for both atoms to be spin-up or spin-down. However, when a weak measurement (ITE) is performed on the system, the measurement result is that atom 1 will only appear in the u path and atom 2 will exit; And the spin of atom 2 can only be observed in the u path, and the spin of atom 1 can only be observed in the d path. The numerical solution obtained here is basically consistent with the analytical solution in Section 2, which theoretically verifies the feasibility of the experiment. When the spin exchange does not occur, according to the analytical solution in Section 2, it can be concluded that the weak value is not directly related to the α, but the measurement success probability is still affected by the α. Therefore, after the delayed selection experiment is introduced into the quantum Cheshire cat effect of a many-particle system, two post-selected States appear in the post-selection, and the two atoms and spins can be positioned on different selections. By switching different post-selected States, the weak values of the two atoms and their spins will produce different results. Since the measurement of the atom and spin positions in the interferometer can be performed independently of the delayed selection process, selecting a direction for post-selection, the effect of delayed selection on the previous measurement can be studied. In order to more directly reflect the impact of delayed selection, we continue to take the weak value of the α=π/4 calculation system.

Compared with the Fig. 3, the Fig. 4 has one more post-selection process, so not only different atoms are distinguished on the data, but also different positions are distinguished in the figure, and the relationship between the normalized coincidence rate of the system and the *t* is shown in the four figures. When the spins of the two atoms are not exchanged, the weak values of the spin and the position follow the same trend. When the spin is exchanged, it is consistent with the weak value in the Fig. 4, which indicates that the delayed selection does have an impact on the system, and the different post-selected states affect the system that has been measured previously. Atoms with different transmittances obtain different post-selected states and are received by the detector. After linear fitting of each point, the numerical solution of the system in delayed selection is calculated according to the expression (11) of the weak value obtained by ITE and the fitting of the experimental results.

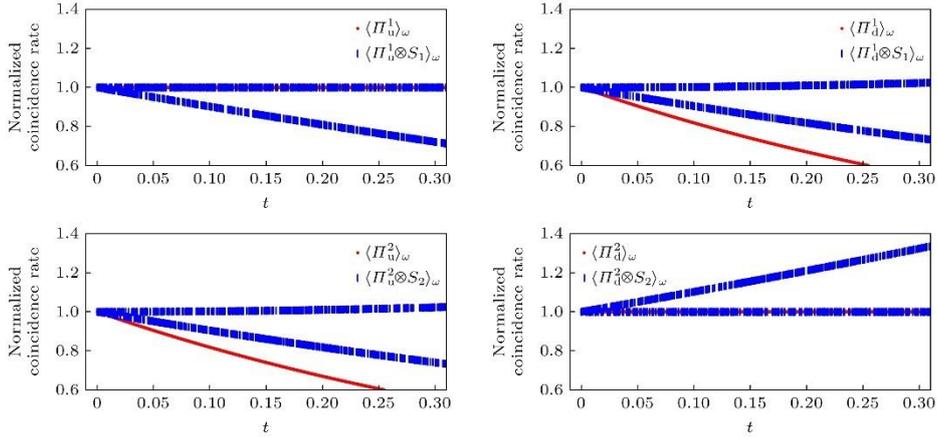

**Figure 4.** Add ITE images to the system after delay selection. The system will obtain two different post selected states, and the normalized coincidence rate of each atom with spin in the evolved system will vary.

$$\langle \Pi_u^1 \rangle_w = 0.00, \quad \langle \Pi_u^2 \rangle_w = 0.83,$$
$$\langle \Pi_d^1 \rangle_w = 0.83, \quad \langle \Pi_d^2 \rangle_w = 0.01. \tag{17}$$

$$\langle \Pi_u^1 \otimes S_1 \rangle_w = 0.22, \quad \langle \Pi_u^2 \otimes S_2 \rangle_w = 0.18,$$
$$\langle \Pi_d^1 \otimes S_1 \rangle_w = 0.17, \quad \langle \Pi_d^2 \otimes S_1 \rangle_w = -0.35. \tag{18}$$

By comparing the numerical and analytical solutions of the system, it can be concluded that when the post-selected state is $|f\rangle$, the spins of the two atoms are exchanged, as expected;

After adding the delayed selection at the post-selection, there is an additional post-selected state in the post-selection process, which is $|f_2\rangle$. When the system is in this post-selected state, the spins of atom 1 and atom 2 are not exchanged, indicating that the measured value corresponds to the properties of atom 1 or atom 2. The analytical solution is consistent with the numerical solution, indicating that the fermion system also satisfies the time symmetry in the multi-particle quantum Cheshire cat effect, that is, the spin exchange between each other occurs randomly, and the post-selection and delayed selection affect the measured value before the measurement.

## 4. Conclusion

In this paper, the influence of the quantum Cheshire cat effect on the system of two spin-1/2 atoms is discussed. By constructing the pre-selected and post-selected states of the atoms, the weak value is obtained by ITE, and the delayed selection in the post-selection makes the spin exchange of the two atoms full of randomness. When the spin exchange occurs, their spin weak values show the characteristics consistent with the tan function, that is, the spin weak value is enlarged or reduced while the spin exchange is realized. The advantage of amplification is that it is easier to obtain more accurate measurements in noisy environments or in environments where the detector is saturated, while the advantage of reduction is to improve the probability of successful measurements. The two can be combined to measure the system multiple times to obtain more complete measurement results. When the two atoms obtain a post-selected state without spin exchange, under the combined action of the quantum Cheshire cat effect and the delayed selection effect, even though the system has evolved, the information of spin and position is still consistent with the information of the system before measurement. In addition, the delayed choice experiment further reveals the physical connotation of the quantum Cheshire cat. Because of the special nature of weak value information, only the appropriate post-selected state can accurately extract useful information. Multi-party communication is realized by attribute exchange, and the anti-interference ability of the signal is improved by weak value amplification, so as to improve the efficiency of signal communication. Based on the above characteristics, we expect that these research results can be applied in the fields of quantum circuits, quantum communication and quantum precision measurement.